\documentclass[10pt, conference, compsocconf]{IEEEtran}

\usepackage[T1]{fontenc}
\usepackage[english]{babel}
\usepackage{rotating,array}
\usepackage[cmex10]{amsmath}
\usepackage{amsfonts}
\usepackage{amssymb}
\usepackage{amscd}
\usepackage{amsthm}
\usepackage{url}
\usepackage{graphicx}
\usepackage{multirow}
\usepackage{subfigure}
\usepackage{booktabs}
\usepackage{textcomp}
\usepackage{stmaryrd}

\newcommand{\be}{\begin{equation}}
\newcommand{\ee}{\end{equation}}

\def\è{\`{e}}
\def\à{\`{a}}
\def\ò{\`{o}}
\def\ù{\`{u}}
\def\ì{\`{i}}
\def\é{\'{e}}

\newtheorem{theorem}{Theorem}

\newtheorem{proposition}[theorem]{Proposition}
\newtheorem{corollary}[theorem]{Corollary}

\newcommand{\Fu}{\textrm{FTL}}

\newcommand{\N}{\mathbb{N}}
\newcommand{\Z}{\mathbb{Z}}

\newcommand{\G}{\mathcal{G}}
\newcommand{\X}{\mathcal{X}}
\newcommand{\F}{\mathcal{F}}
\newcommand{\W}{\mathcal{W}}

\newcommand{\Last}{\mathcal{L}}
\newcommand{\Lt}{\mathcal{L}_t}

\newcommand{\means}[1]{\hbox{$ [\kern -.4em [\, {#1}\, ]\kern -.4em]$}}

\DeclareMathOperator{\AG}{\mathcal{A}\mathcal{G}}
\DeclareMathOperator{\AU}{\mathcal{A}\mathcal{U}}
\DeclareMathOperator{\Soon}{\mathcal{S}\textit{o}\textit{o}\textit{n}}
\DeclareMathOperator{\U}{\mathcal{U}}

\begin{document}
\IEEEoverridecommandlockouts

\title{Fuzzy Time in LTL}

\author{\IEEEauthorblockN{Achille Frigeri}
\IEEEauthorblockA{Politecnico di Milano\\
p.zza L. da Vinci, 32\\
20133 - Milano, Italy\\
 \tt{achille.frigeri@gmail.com}}
\and
\IEEEauthorblockN{Liliana Pasquale}
\IEEEauthorblockA{Lero, Irish Software Eng. Research Centre\\
 University of Limerick\\
Limerick, Ireland\\
 \tt{liliana.pasquale@lero.ie}}
\and
\IEEEauthorblockN{Paola Spoletini}
\IEEEauthorblockA{Universit\`a dell'Insubria\\
via Ravasi, 2\\
21100 - Varese, Italy\\
\tt{paola.spoletini@uninsubria.it}}}

\maketitle

\begin{abstract}
  In the last years, the adoption of active systems has increased in
  many fields of computer science, such as databases, sensor networks,
  and software engineering. These systems are able
  to automatically react to events, by collecting information from
  outside and internally generating new events. However, the
  collection of data is often hampered by uncertainty and vagueness
  that can arise from the imprecision of the monitoring
  infrastructure, unreliable data sources, and networks. The decision
  making mechanism used to produce a reaction is also imprecise, and
  cannot be evaluated in a crisp way. It depends on the evaluation of
  vague temporal constraints, which are expressed on the
  collected data by humans. Despite fuzzy logic has been mainly conceived as a
  mathematical abstraction to express vagueness, no attempt has been
  made to fuzzify the temporal modalities. Existing fuzzy languages do
  not allow us to represent temporal properties, such as
  ``almost always'' and ``soon''. Indeed, the semantics of existing
  fuzzy temporal operators is based on the idea of replacing classical
  connectives or propositions with their fuzzy counterparts.  To
  overcome these limitations, we propose a temporal framework, FTL
  (Fuzzy-time Temporal Logic), to express vagueness on time. This
  framework formally defines a set of fuzzy temporal modalities, which
  can be customized by choosing a specific semantics for the
  connectives. The semantics of the language is sound, and the
  introduced modalities respect a set of expected mutual
  relations. We also prove that under the assumption that all events
  are crisp, FTL reduces to LTL. Finally, for some of the possible
  fuzzy interpretations of the connectives, we identify adequate
  sets of temporal operators, from which it is possible to derive all
  the others.
\end{abstract}



\IEEEpeerreviewmaketitle

\section{Introduction}
\label{sec:intro}


In the last years, the adoption of active systems has increased in many fields of computer science. Active systems must automatically react to achieve or maintain their requirements,
depending on the information collected from the surrounding
environment. Examples of such systems are
active databases \cite{activeDB}, active sensor networks
\cite{activeSN}, and smart grids~\cite{SmartGrids}. For instance,
smart grids may need to adjust the workload on the appliances (e.g.,
fridge, oven) available in a building to optimize energy
consumption and costs.

Event-driven architectures~\cite{EventBasedSys} are a common architectural
paradigm to design active systems. This paradigm is based on the idea
that the actions the system will perform are generated as a reaction to the events
occurred inside and outside the system. In many
cases, providing such active functionality requires to materialize the
occurrence of other relevant events, according to a set of inference rules. These rules are
generally defined by domain experts, and are formalized by designers.
Domain experts must provide the set of basic events to be collected, which serve as
input to the rules, their inter-relationships, and the parameters of
the events for determining a new event materialization.

However, the collection of data is often
hampered by uncertainty and vagueness that can arise from the
imprecision of the monitoring infrastructure, unreliable data
sources, and networks. The inference rules that are used to produce
a reaction are also imprecise. They often depend on the evaluation
of untimed or temporal properties that
are vague, since they are expressed by
humans, and, for this reason, cannot be assessed in a crisp
way. For example, a smart
grid must satisfy the following property: ``all appliances must be
available almost always''. This rule is vague since the concept of
availability cannot be precisely assessed, because it may depend on
the perception of the customers. The
temporal period (``almost always''), during which the availability
property must be satisfied, is vague as well. For these reasons, it
becomes fundamental to identify a suitable
formalism to represent vague properties as suitable untimed or temporal formulae.

Fuzzy logic has been conceived as a mathematical abstraction to
express vagueness in the satisfaction of formulae.
While the propositional fuzzy logic has been deeply
investigated, the fuzzy version of the temporal modalities has been
often neglected. Few attempts~\cite{Thiele93,Lamine00,Dutta88,Dubois89,Moon04} to manage time have been made, but all these approaches just focus on the
uncertainty of the information and do not take into account the truth
degree of temporal expressions. The semantics of existing fuzzy temporal
operators is based on the idea of replacing classical connectives or
propositions with their fuzzy counterparts. Existing fuzzy languages do not allow us to represent
additional temporal properties, such as ``almost always'',
``soon''. This kind of modalities may be useful when we need to specify situations when a formula is slightly satisfied, since an event
happens a little bit later than expected, when a property is always satisfied except for a small set of time instants, or
a property is maintained for a time interval which is slightly smaller
than the one requested.

To overcome these limitations, we propose a temporal framework, FTL (Fuzzy-time  Temporal Logic), to express vagueness on time. This framework formally defines a set of fuzzy temporal modalities, which can be customized by choosing a specific semantics for the connectives.
The semantics of the language is sound, and the introduced
modalities respect a set of expected mutual relations. We also
prove that under the assumption that all events are crisp, FTL
reduces to LTL. Finally, for some of the possible fuzzy
interpretations of the connectives, we identify an adequate set of
temporal operators, from which it is possible to derive all the others.



The paper is organized as follows. Section \ref{sec:motivations}
discusses some related work. Section \ref{sec:fuzzy} provides some background knowledge about fuzzy logic and points out its differences w.r.t. probability theory.
Section \ref{sec:language} presents the \Fu\ framework, by
illustrating some interesting properties of the operators it
introduces. Section \ref{sec:reductions} identifies an adequate set of
connectives for some classical interpretations of the connectives. Note that proofs of propositions are given in a sketch form, and minor details are left to the reader.
Section \ref{sec:example} provides some example of possible FTL
specifications in the context of smart grids, and Section
\ref{sec:conclusions} concludes the paper.

\section{Related Work}
\label{sec:motivations}


In computer science, fuzzy logic has been mainly used to represent the
uncertainty due to the unpredictability of the environment or the
imprecision of the measurements. Many
attempts~\cite{Lamine00,Thiele93} have been made to use fuzzy logic to
monitor the satisfaction of temporal properties of the system and/or
the environment. For example, Lamine and Kabanza~\cite{Lamine00} add,
for each classic temporal operator (e.g., always, eventually, until,
etc.), a corresponding fuzzy temporal one. This operator keeps the
same semantics of its crisp counterpart, with the only difference that
the Boolean connectives (not, and, or) are replaced with the
corresponding operations in the Zadeh interpretation (see operations
associated respectively with negation, t-norm and t-conorm in
Table~\ref{tab:norme}).  The authors evaluate a fuzzy proposition over
a history (i.e., a sequence of states) and associate a weight with the
evaluation made at each state.  The weights and the extent to which
the history is needed to evaluate a proposition are defined
empirically, depending on the application and the properties expressed
by the proposition itself. Similarly, Thiele and
Kalenka~\cite{Thiele93} define a fuzzy ``interpretation'' of the
traditional temporal operators. They also introduce proper fuzzy
temporal operators to represent the short or long time distance in
which a specific property must be satisfied (in the future or in the
past). Despite the aforementioned approaches are a first step towards
the fuzzyfication of time, they do not associate a specific fuzzy
semantics with the temporal modalities. Instead, temporal modalities
have a fuzzy semantics only depending on the interpretation given to
their (sub-)argument, which is an untimed fuzzy formula.


Other works~\cite{Dutta88,Dubois89,Moon04} have a slightly different
objective. They use fuzzy temporal logic to express uncertainty about
the time in which some specific events may occur and the temporal
relationships among events and states. Dutta~\cite{Dutta88} defines
the occurrence of an event as the possibility of its occurrence in any
time interval. This way the authors can evaluate a set of temporal
relations between a pair of events: if an event precedes/follows another one,
the degree an event overlaps another one, or whether an event
immediately follows another one. Similarly, Dubois and
Prade~\cite{Dubois89} represent dates as a possibility
distribution. Hence, it is possible to express different situations:
whether a date is precisely known or not (i.e., it is within an
interval), whether a date is fuzzily known (i.e., the interval
boundaries that contain the date are not clearly known), or whether a
date is attached to an event that may not occur. From this
representation the authors use fuzzy sets to represent time points
that are possibly/necessarily after or before a date, and use fuzzy
comparators to express relations between time instants. Finally, Moon
et al.~\cite{Moon04} do not consider uncertainty on the time instants,
but fuzzify temporal events and states and define an order relation
among events and states, represented as a directed graph.


In requirements engineering fuzzy logic has been adopted to perform tradeoff
analysis~\cite{ICSE96} among conflicting functional requirements. In
particular, aggregation functions are used to combine correlated
requirements into high-level ones. Fuzzy logic has been also exploited
to express uncertain requirements~\cite{SpecifyFuzzy,RELAX,re}.
Liu et al.~\cite{SpecifyFuzzy} introduce a methodology to elicit
non-functional requirements through fuzzy membership functions that
allow one to represent the uncertainty about the human perception.
RELAX~\cite{RELAX} is a notation to express
uncertain requirements, whose assessment is affected by the
imprecision of measurement.
Finally, FLAGS~\cite{re} extends traditional LTL by
adding new operators to represent transient/small violations
in the temporal domain. Its main purpose is providing a notion of
satisfaction level of requirements in the temporal domain. In particular, the authors
use this approach to tolerate small deviations of the satisfaction of
the requirements during or within a temporal interval. Despite the
purpose of FLAGS is similar to our approach, the syntax and the
semantics of the FLAGS language are not formally described, and the
relations among temporal operators are not even provided.

\section{Background}
\label{sec:fuzzy}

This section provides a general definition of fuzzy logic, and
points out the differences between a fuzzy and a
probabilistic approach for the evaluation of temporal
properties. Finally, the section introduces the formalism proposed in
\cite{Lamine00} and discusses its limitations.

\subsection{General formalization of fuzzy logic}

The term \lq\lq fuzzy\rq\rq\ has been explicitly used for the first time in Zadeh's seminal
work~\cite{zadeh} about fuzzy sets, where he presented the theory of
classes with unsharp boundaries. In this work, the logical formalism
of fuzzy sets shares the same syntax of Propositional Logic (PL),
but its formulae may have a truth value comprised between 0 and
1. Conjunction and disjunction are interpreted as $\min$
and $\max$ operations, respectively.

As Zadeh pointed out~\cite{Zadeh94}, two main directions in fuzzy logic have to be distinguished. In a broad sense, fuzzy logic has been used
to support fuzzy control and to express the vagueness of natural languages,
without demonstrating its formal properties. In a narrow sense, \lq\lq fuzzy logic is a logical system which is an extension of multivalued logic and is intended to serve as a logic of approximate reasoning\rq\rq.
In this paper,
we use the term \lq \lq fuzzy logic\rq\rq\ to refer both to the Zadeh
Logic~\cite{zadeh} (which in computer science it is often called
\lq\lq Fuzzy Logic\rq\rq) and each continuous t-norm fuzzy logic~\cite{Hajek98}. Despite the Zadeh Logic has been
heavily applied in soft computing, it has no strong logical
characterization. Instead, for t-norm fuzzy logics, it is often
possible to provide an axiomatization and some completeness results.


We conceive a fuzzy logic as a  many-valued
logic~\cite{Lukazz}, whose formulae may have a truth value comprised
between 0 and 1 and the
semantics of the connectives satisfies some monotonicity laws.
The semantics of a fuzzy logic must also be \emph{coherent} with PL,
which means that fuzzy logic and PL must share the same syntax, fuzzy logic must reduce
to PL when all predicates assume value 0 or 1, and conjunction and
disjunction must be commutative and associative connectives.
The semantics of the conjunction ($\wedge$), disjunction ($\vee$),
negation ($\neg$), and implication ($\Rightarrow$), is inferred by
considering respectively a continuous t-norm
($\otimes$)~\cite{Klement}, its associated t-conorm ($\oplus$), a
negation function ($\ominus$), and an implication function
($\varogreaterthan$). In the case of a t-norm fuzzy logic, the negation is the pseudo-complement (i.e., $\ominus \alpha
=\max\{\beta\in [0,1]\mid \alpha \otimes \beta=0\}$), while the
implication function becomes the residuum of the t-norm (i.e., $\alpha
\varogreaterthan\beta=\max\{\gamma\in [0,1]\mid \alpha \otimes
\gamma\leq \gamma\}$). In the rest of the paper we will refer to these
functions as the interpretation of connectives. Note also that the
family of (continuous) t-norm fuzzy logics is infinite, as
demonstrated by the infinite class of Dubois-Prade \cite{Dubois82} and
the Yager \cite{Yager} t-norms and t-conorms.

Table~\ref{tab:prop}
summarizes some useful properties of the connectives of a fuzzy logic, while
Table~\ref{tab:norme} provides the interpretation of these connectives
for the Zadeh Logic and three other well-known
t-norm fuzzy logics.

\begin{table*}[Ht]
\caption{Properties of the interpretations of connectives.}\label{tab:prop}
\centering
\begin{tabular}{c|c|c|c|c}
\toprule
 & boundary value & commutativity & associativity & monotonicity \\
  \midrule
negation & \begin{tabular}{l}$\ominus 0=1$\\$\ominus 1=0$\end{tabular} & - & - & $\alpha\leq \beta\Rightarrow \ominus \alpha \geq \ominus \beta$ \\
  \midrule
t-norm & \begin{tabular}{l}$\alpha\otimes 0=0$\\$\alpha\otimes 1=\alpha$\end{tabular} & yes & yes & \begin{tabular}{c}$\beta\geq \gamma\Rightarrow \alpha\otimes \beta\geq\alpha\otimes\gamma$\\ $\alpha\otimes\beta\leq\alpha$\end{tabular}\\
  \midrule
t-conorm & \begin{tabular}{l}$\alpha\oplus 0=\alpha$\\$\alpha\oplus 1=1$\end{tabular} & yes & yes & \begin{tabular}{c}$\beta\geq \gamma\Rightarrow \alpha\oplus \beta\geq\alpha\oplus\gamma$\\ $\alpha\oplus \beta\geq \alpha$\end{tabular} \\
  \midrule
implication &
\begin{tabular}{c} $1\varogreaterthan \beta=\beta$\\
$0\varogreaterthan\beta=\alpha\varogreaterthan 1=1$\\ $\alpha\varogreaterthan 0=\ominus \alpha$\end{tabular}  & no & no &
\begin{tabular}{c}
$\alpha\leq \beta\Rightarrow \alpha\varogreaterthan \gamma\geq\beta\varogreaterthan\gamma$\\
$\beta\leq \gamma\Rightarrow \alpha\varogreaterthan \beta\leq\alpha\varogreaterthan\gamma$\\
$\alpha\varogreaterthan\beta\geq \max\{\ominus\alpha,\beta\}$\\
\end{tabular}\\
\bottomrule
\end{tabular}
\end{table*}

\begin{table*}[Ht]
\caption{Some interpretation for connectives.}\label{tab:norme}
 \centering
\begin{tabular}{c|c|c|c|c}
      \toprule
 & Zadeh~\cite{zadeh} & G\"{o}del-Dummett~\cite{Dummett} & {\L}ukasiewicz~\cite{Lukazz} & Product~\cite{Hajek}\\
  \midrule
$\ominus\alpha$ & $1-\alpha$ & $\left\{
                           \begin{array}{ll}
                             1, & \hbox{$\alpha=0$} \\
                             0, & \hbox{$\alpha>0$}
                           \end{array}
                         \right.$
 & $1-\alpha$ & $\left\{
                           \begin{array}{ll}
                             1, & \hbox{$\alpha=0$} \\
                             0, & \hbox{$\alpha>0$}
                           \end{array}
                         \right.$ \\
  \midrule
$\alpha\otimes\beta$ & $\min\{\alpha,\beta\}$ & $\min\{\alpha,\beta\}$ & $\max\{\alpha+\beta-1,0\}$ & $\alpha\cdot\beta$  \\
  \midrule
$\alpha\oplus\beta$ & $\max\{\alpha,\beta\}$ & $\max\{\alpha,\beta\}$ & $\min\{\alpha+\beta,1\}$ & $\alpha+\beta-\alpha\cdot\beta$  \\
  \midrule
$\alpha\varogreaterthan\beta$ & $\max\{1-\alpha,\beta\}$ & $\left\{
                           \begin{array}{ll}
                             1, & \hbox{$\alpha\leq \beta$} \\
                             \beta, & \hbox{$\alpha>\beta$}
                           \end{array}
                         \right.$ & $\min\{1-\alpha+\beta,1\}$ & $\left\{
                           \begin{array}{ll}
                             1, & \hbox{$\alpha\leq \beta$} \\
                             \beta/\alpha, & \hbox{$\alpha>\beta$}
                           \end{array}
                         \right.$\\
   \bottomrule
\end{tabular}
\end{table*}

Once identified an interpretation of the connectives, the evaluation of a (fuzzy)
formula can be represented as a function $v_i$ from the set of
well-formed formulae to $[0,1]$, which extends the interpretation
$i:AP\rightarrow [0,1]$ used to evaluate an atomic proposition
in $AP$.

The following proposition describes some well-known properties of t-norms and t-conorms.

\begin{proposition}\label{p:limiti}
Let $\otimes$ be a t-norm and $\oplus$ be a t-conorm, $\alpha,\beta\in
[0,1]$, and $d^+,d^{\times}:[0,1]^2\rightarrow \{0,1\}$ be the drastic
sum and the drastic product defined respectively by:
\[
\begin{array}{l}
d^+(\alpha,\beta)=1 \Leftrightarrow \alpha+\beta>0,\\
d^{\times}(\alpha,\beta)=1 \Leftrightarrow \alpha\cdot \beta=1.
\end{array}
\]
Then
\[
\begin{split}
\max\{\alpha,\beta\}&\leq \alpha\oplus \beta \leq d^+(\alpha,\beta),\\
d^{\times}(\alpha,\beta) &\leq \alpha\otimes \beta \leq \min\{\alpha,\beta\}.
\end{split}
\]
\end{proposition}

For a continuous t-norm it is possible to define two connectives
called lattice (or weak) conjunction ($\wedge^w$), and lattice
disjunction ($\vee^w$). The semantics of these connectives is
given by:
\[
\begin{split}
&p\wedge^w q=p\wedge(p\Rightarrow q),\\
&p\vee^w q=((p\Rightarrow q) \Rightarrow q) \wedge^w  ((q\Rightarrow p) \Rightarrow p).
\end{split}
\]
Nevertheless, they reduce respectively to the $\max$ and $\min$ operations, as stated in the following well-known proposition.

\begin{proposition}\label{p:max}
Let $p_\alpha,p_\beta\in AP$ such that $i(p_\alpha)=\alpha$, and $i(p_\beta)=\beta$, then, for each continuous t-norm:
\[
\begin{array}{l}
    v_i(p_\alpha\wedge^w p_\beta)=\max\{\alpha,\beta\},\\
    v^i(p_\alpha\vee^w p_\beta)=\min\{\alpha,\beta\}.
\end{array}
\]
\end{proposition}

\subsection{Fuzzy Logic and Probability}

Fuzzy logic and probability have been usually
conceived as similar disciplines. However, the nature of fuzzy logic
and probability are totally different both on the ontological and
epistemological level. These disciplines
deal with two different topics. Probability focuses on observable
events whose occurrence is uncertain, while fuzzy logic deals with
vague events that cannot be clearly assessed.

For example, the
statement \lq\lq tomorrow there will be a power outage\rq\rq\ is
uncertain, since it is not possible to know the truth value of the
formula. However, by applying the probability
theory (e.g., by
analyzing the frequency of power outages during the last month), it is possible to
state that, for example, the probability that the aforementioned
statement will be true is $3.8\%$.
Still, when a direct observation can be performed (i.e.,
tomorrow), it is possible to assess whether an outage took place or
not and, indeed, the probability value can collapse either to $0$ or $1$.

Instead, the
statement ``tomorrow the number of power outages will be low'' is not
tractable from a probabilistic point of view, because the nature of the event itself is
not clearly measurable, since the concept of ``low'' has not been
defined in a observable way. In this case, we are not facing the
problem of uncertainty of an event, but the vagueness of its
definition. Indeed, assigning the truth degree of $0.038$ to the aforementioned
statement means that
tomorrow the smart grid will face a ``high number of outages''. Even a
direct observation of the number of outages will not cause this value to collapse to $0$ or $1$.

\subsection{Fuzzy Linear-time Temporal Logic}

This section briefly describes FLTL (Fuzzy Linear-time Temporal
Logic)~\cite{Lamine00}, which is an extension of Zadeh Logic with temporal operators. FLTL has the same syntax of LTL. In particular, let
$\Phi$ be the set of well formed formulae and $AP$ the set of
propositional letters, then $\varphi\in \Phi$ if and only if
\[
\varphi:=p\mid \neg \varphi \mid \varphi \wedge \varphi \mid \mathbf{X} \varphi \mid \mathbf{G} \varphi \mid \varphi \mathbf{U} \varphi,
\]
where $p\in AP$.
The semantics of a formula $\phi\in \Phi$ is defined w.r.t. a
linear time structure $\pi_{\sigma}=(S, w_0, w, L)$, where $S$ is a set of
states, $w_0$ is the initial state,
$w\in w_0 S^\omega$ is an \emph{infinite path},
and $L: S \to [0,1]^{AP}$ is a \emph{fuzzy labeling function}.
The evaluation $v(\varphi,w^i)$ of a formula $\varphi\in \Phi$ along the path $w$ from the $i$-th instant is a real number in $[0,1]$ recursively defined by:
\begin{equation*}
\begin{array}{l}
v(p,w^i)=L(w_i)(p), \\
v(\neg \varphi,w^i)=1-v(\varphi,w^i),\\
v(\varphi\wedge \psi,w^i)=\min\{v(\varphi,w^i),v(\psi,w^i)\},\\
v(\mathbf{X} \varphi,w^i)=v(\varphi,w^{i+1}),\\
v(\mathbf{G} \varphi,w^i)=\min\{v(\varphi,w^{i}),v(\mathbf{G} \varphi,w^{i+1})\},\\
v(\varphi \mathbf{U} \psi,w^i)=\\
\qquad\qquad\max\{v(\psi,w^{i}),\min\{v(\varphi,w^i),v(\varphi \mathbf{U} \psi,w^{i+1})\}\}.
\end{array}
\end{equation*}
It is easy to see that FLTL extends LTL in the sense that if for all $s\in S$ and $p\in AP$ is $L(s)(p)\in \{0,1\}$, then $v(\varphi,w^i)=1\Leftrightarrow w^i\models \varphi$.

Note that FLTL cannot
represent the vagueness in the temporal dimension. Fuzzyfication
just addresses Boolean connectives and keeps a crisp semantics for the time
(always/never). For example, when we evaluate the ``globally''
(always) operator, it
may not be suitable to consider the minimum truth value 
encountered. For instance, this semantics does not allow us to
tolerate transient violations that take place for a few number of
times compared to a long time interval.  For example, if we want to
assess the truth of the statement ``this week no power outage
happened'', we must consider that even one power outage is enough to
negatively affect the truth value of this formula, and we cannot
tolerate a few power outages. Furthermore, even if
this semantics allows us to express statements about the future, such
as ``tomorrow power outages will take place'', we cannot express
statements, such as ``soon a power outage will happen''.

For these reasons, the language we propose in this paper, although partially inspired by FLTL, introduces a completely new approach to
the fuzzifycation of the temporal domain.

\section{\Fu: Fuzzy-time Temporal Logic}
\label{sec:language}

In this section we describe the syntax and
semantics of \Fu, which is our fuzzy-time temporal logic.

\subsection{Syntax}

\Fu\ extends LTL in order to deal with fuzzyness on time. Let $AP$ be a numerable set of atomic propositions, $\neg,\
\wedge,\ \vee, \Rightarrow$ be the (fuzzy) connectives, and
$O$ and $T$ be the sets of unary and binary (fuzzy) temporal
modalities. Then, $\varphi$ belongs to the set $\Phi$ of
\emph{well-formed \Fu\ formulae} (from now on, simply \emph{formulae})
if it is defined as follows:
\[
  \varphi := p \mid \neg~\varphi \mid \varphi~\sim~\varphi \mid \mathcal{O}\varphi \mid \varphi\mathcal{T}\varphi,
\]
where $p \in AP$, $\sim$ is a binary connective, $\mathcal{O}\in O$,
and $\mathcal{T}\in T$. As unary operators we consider $\X$ (next),
$\Soon$ (soon), $\F$ (eventually), $\F_t$ (eventually in the next $t$
instants), $\G$ (always), $\G_t$ (always in the next $t$ instants),
$\AG$ (almost always), $\AG_t$ (almost always in the next $t$ instants), $\Last_t$ (lasts $t$ instants), $\W_t$ (within $t$ instants), where $t\in \N$. Binary operators are $\U$ (until), $\U_t$ (bounded until), $\AU$ (almost until), and $\AU_t$ (bounded almost until).
We admit the use of $\X^j(\cdot)$ as a shorthand for
$j$ applications of $\X$. For example, $\X^2(\cdot) \equiv
\X(\X(\cdot))$. Conventionally we also set $\X^0 \varphi\equiv \varphi$.
From now on, operators $\Soon$, $\AG$, $\AG_t$, $\Last_t$, $\W_t$, $\AU$,
and $\AU_t$ will be indicated as \lq \lq almost\rq \rq\ operators.


\subsection{Semantics}

The semantics of a formula $\varphi$ is defined w.r.t.
a linear time structure $(S,s_0,\pi, L)$, where $S$ is the set of
states, $s_0$ is the initial state, $\pi$ is an infinite path $\pi =
s_0s_1\dots\in S^{\omega}$, and $L: S \to [0,1]^{AP}$ is the
\emph{(fuzzy) labeling function} that assigns to each state an
evaluation for each atomic proposition in $AP$. $\pi^i$ indicates the
suffix of $\pi$, by starting from the $i$-th
position and $s^i$ is the first state of $\pi^i$.
Besides, we adopt an \emph{avoiding function} $\eta:\Z\rightarrow
[0,1]$. We assume that $\eta(i)=1$ for all $i\leq 0$, and $n_{\eta}\in\N$ exists such
that $\eta$ is strictly decreasing in $\{0,\dots,n_{\eta}\}$ and $\eta(n')=0$
for all $n'\geq n_{\eta}$.
Function $\eta$ expresses the penalization assigned to the number of events we want
to ignore in evaluating the truth degree of a formula that
contains an \lq\lq almost\rq\rq\ operator. For
example, we interpret the formula \lq\lq almost always $p$\rq\rq\ as
\lq\lq always $p$ except for a small number of cases\rq\rq, and we
penalize the evaluation of the formula according to the number of avoided events. Hence, the evaluation of a formula that contains the operator
$\AG$ realizes a tradeoff between the number of avoided
events, and the penalization assigned to this number.

Since we are dealing with a multi-valued logic, it makes no sense to
define a crisp satisfiability relation. Instead, to define
the semantics of a formula $\varphi$ along a path, we express a \emph{fuzzy satisfiability relation} as $\models~\subseteq~S^{\omega}\times F\times
[0,1]$, where $(\pi\models \varphi)=\nu\in [0,1]$ means that the
truth degree of $\varphi$ along $\pi$ is $\nu$.
We say that two formulae $\varphi$ and $\psi$ in $\Phi$ are \emph{logically equivalent}, in symbols $\varphi\equiv\psi$, if, and only if, $(\pi\models \varphi)=(\pi\models\psi)$ for each linear time structure, and for each avoiding function.

The truth degree of a formula is defined, as usual, recursively on its structure. Let $p\in AP$ and $\pi^i$ be a path, then:
\begin{equation*}
\begin{array}{l}
(\pi^i\models p)=L(s^i)(p), \\
(\pi^i\models \neg\varphi)=\ominus (\pi^i\models \varphi), \\
(\pi^i\models \varphi\wedge\psi)=(\pi^i\models \varphi)\otimes (\pi^i\models \psi), \\
(\pi^i\models \varphi\vee\psi)=(\pi^i\models\varphi) \oplus (\pi^i\models \psi), \\
(\pi^i\models \varphi\Rightarrow\psi)=(\pi^i\models \varphi)\varogreaterthan (\pi^i\models \psi),
\end{array}
\end{equation*}
where $p\in AP$, $i\in \N$, and
$\ominus,\otimes,\oplus,\varogreaterthan$ are the operations, between
real numbers, defining the chosen semantics of the connectives
($\neg,\ \wedge,\ \vee, \Rightarrow$).

We are now able to introduce the semantics of FTL temporal operators.

\subsubsection*{\bf Next}
Operator \lq\lq next\rq\rq\ ($\X$) has the same semantics of its corresponding LTL operator $\mathbf{X}$:
\[
(\pi^i\models \X\varphi)=(\pi^{i+1}\models\varphi).
\]

\subsubsection*{\bf Soon}
Operator \lq\lq soon\rq\rq\ ($\Soon$) extends the semantics of the \lq\lq next\rq\rq\ operator, by
tolerating at most $n_{\eta}$ time instants of delay. In other words,
the greater the
number of tolerated instants, the greater the penalization will be.
\[
(\pi^i\models \Soon\varphi)=\bigoplus_{j=i+1}^{i+n_\eta} (\pi^{j}\models\varphi)\cdot \eta(j-i-1).
\]
\begin{proposition}\label{p:soon}
From the monotonicity of the t-conorm $\oplus$ (see Table
\ref{tab:prop}) it naturally follows that
\[
(\pi^i\models \X \varphi) \leq (\pi^i\models \Soon \varphi).
\]
\end{proposition}

\subsubsection*{\bf Eventually}
Operator \lq\lq eventually\rq\rq\ ($\F$) and its
bounded version ($\F_t$) also maintain the same semantics of their corresponding
LTL operator $\mathbf{F}$. Namely,
\[
\begin{split}
(\pi^i\models \F_t\varphi)&=\bigoplus_{j= i}^{i+t} (\pi^{j}\models\varphi),\\
(\pi^i\models \F\varphi)&=\bigoplus_{j\geq i} (\pi^{j}\models\varphi)=\lim_{t\rightarrow +\infty} (\pi^i\models \F_t\varphi).
\end{split}
\]
First, observe that for $\F_t$ the equivalences $\F_0\varphi\equiv
\varphi$ and $\F_t\varphi\equiv \varphi\vee \X\F_{t-1}\varphi$ hold, for $t\geq 0$.
The semantics of $\F$ requires a passage to the limit, whose existence is ensured by the fact that the sequence $(\pi^i\models \F_t\varphi)_{t\in \N}$ is increasing, as the t-conorm $\oplus$ is monotonic. These facts are summarized
in the following proposition.

\begin{proposition}\label{p:eventually}
For all $\varphi\in F$ and $t\leq t'$:
\[
(\pi^i\models \varphi) \leq (\pi^i\models \F_{t} \varphi) \leq (\pi^i\models \F_{t'} \varphi) \leq (\pi^i\models \F \varphi).
\]
\end{proposition}

\subsubsection*{\bf Within}
Operator \lq\lq within\rq\rq\  ($\W_t$) is inherently bounded, and its semantics is defined by
\[
(\pi^i\models \W_t \varphi)=\bigoplus_{j=i}^{i+t+n_{\eta}-1} (\pi^j\models\varphi)\cdot \eta(j-t-i).
\]
Formula $\W_t p$ states that subformula $p$ is supposed to hold
in at least one of the next $t$ instant or, possibly, in the next
$t+n_{\eta}$. In the last case we apply a penalization for each
instant after the $t$-th.
\begin{proposition}\label{p:within}
The semantics of operator $\W_t$ can be expressed by only using
operators $\X$ and $\Soon$. More formally, for all $\varphi\in F$ and $t\in\N^0$:
\[
\W_t\varphi\equiv \F_t \varphi \vee \X^{t+1}\Soon\varphi,
\]
and
\[
\W_0\varphi\equiv \Soon \varphi.
\]
\end{proposition}

\begin{corollary}\label{c:within}
For all $\varphi\in F$ and $t\in\N$
\[
\begin{split}
&(\pi^i\models \W_t\varphi)\geq(\pi^i\models \F_t\varphi),\\
&\lim_{t\rightarrow +\infty} (\pi^i\models \W_t\varphi)=(\pi^i\models \F\varphi).
\end{split}
\]
\end{corollary}
\begin{IEEEproof}
The first property follows immediately from the previous
proposition. For the second property, observe that $(\pi^i\models \F_{n_\eta})\varphi\geq(\pi^i\models \Soon \varphi)$, and then actually
\[
(\pi^i\models \F_{t+n_\eta}\varphi)\geq(\pi^i\models \W_t\varphi)\geq(\pi^i\models \F_t\varphi),\\
\]
and applying the squeeze theorem we have the thesis.
\end{IEEEproof}

\subsubsection*{\bf Always}
Operator ``always'' ($\G$) and its bounded version ($\G_t$)
extend the semantics of their corresponding LTL operator $\mathbf{G}$. Namely,
\[
\begin{split}
(\pi^i\models \G_t\varphi)&=\bigotimes_{j= i}^{i+t} (\pi^{j}\models\varphi),\\
(\pi^i\models \G\varphi)&=\bigotimes_{j\geq i} (\pi^{j}\models\varphi)=\lim_{t\rightarrow +\infty} (\pi^i\models \G_t\varphi).
\end{split}
\]
As for $\F_t$, observe that for $\G_t$ the equivalences
$\G_0\varphi\equiv \varphi$ and $\G_t\varphi\equiv \varphi\wedge
\X\G_{t-1}\varphi$ hold, for $t\geq 0$.
Similarly to $\F$, the semantics of $\G$ also requires a passage to
the limit, whose existence is ensured by the fact that the sequence $(\pi^i\models \G_t\varphi)_{t\in \N}$ is decreasing, as the t-norm $\otimes$ is monotonic (see Table \ref{tab:prop}). These facts are summarized
in the following proposition.

\begin{proposition}\label{p:always}
For all $\varphi\in F$ and $t\leq t'$:
\[
\begin{split}
(\pi^i\models \G\varphi)&\leq (\pi^i\models \G_t\varphi)\leq (\pi^i\models \G_{t'}\varphi)\\
&\leq (\pi^i\models \G_1\varphi)=(\pi^i\models \varphi\wedge\X\varphi)\\
&\leq (\pi^i\models \G_0\varphi)= (\pi^i\models \varphi).
\end{split}
\]
\end{proposition}

From propositions \ref{p:soon}, \ref{p:within} and \ref{p:always}, we can immediately obtain the following corollary.

\begin{corollary}\label{c:always}
For all $\varphi\in F$ and $t,t'\in \N$:
\[
\begin{split}
(\pi^i\models \G\varphi)& \leq (\pi^i\models \F\varphi),\\
(\pi^i\models \G_t\varphi)& \leq (\pi^i\models \F_{t'}\varphi),\\
(\pi^i\models \G_t\varphi)& \leq (\pi^i\models \W_{t'}\varphi).
\end{split}
\]
\end{corollary}

\subsubsection*{\bf Almost always}

Operator \lq\lq almost always\rq\rq\ ($\AG$) and its bounded version
($\AG_t$) allow us to evaluate a property over the path $\pi^i$, by
avoiding at most $n_\eta$ evaluations of this property, and, at the same time, introducing a penalization for each avoided case.
Let $I_t$ be the initial segment
of $\N$ of length $t+1$, i.e., $I_t=\{0,1,\dots,t\}$, and
$\mathcal{P}^k(I_t)$ the set of subsets of $I_t$ of cardinality $k$,
then
\[
\begin{split}
(\pi^i\models \AG_t \varphi)&=\max_{j\in I_t}\max_{H\in \mathcal{P}^{t-j}(I_t)}\bigotimes_{h\in H}(\pi^{i+h}\models \varphi)\cdot \eta(j),\\
(\pi^i\models \AG \varphi)&=\lim_{t\rightarrow +\infty}(\pi^i\models \AG_t \varphi).\\
\end{split}
\]
As we will see later, the sequence $(\pi^i\models \AG_t
\varphi)_{t\in\N}$ is not monotonic. Nevertheless, we can still prove that the semantics of $\AG$ is well-defined.

\begin{proposition}\label{p:def-almostalways}
Given $\varphi\in F$, it is possible to recursively define $n$ propositional letters $p_0,\dots,p_{n-1}$, such that
\begin{equation}\label{e:ag}
(\pi^i\models \AG \varphi)=\max_{j\leq n_\eta-1}\left\{\G p_j\cdot \eta(j)\right\}.
\end{equation}
\end{proposition}
\begin{IEEEproof}
Let define $p_0$ as:
\[
\forall i\in \N,\ (\pi^i\models p_0)=(\pi^i\models \varphi).
\]
Then, for all $0<m\leq n_\eta$, we recursively obtain $p_m$ from
$p_{m-1}$ in the following way. Let $h_m$ be the minimum in $\N\cup
\{\infty\}$, such that for all $k\in \N$, $(\pi^h\models
p_m)\leq(\pi^k\models p_m)$. Then, let set
\[
\left\{
  \begin{array}{ll}
    (\pi^j\models p_m)=(\pi^j\models p_{m-1}), & \hbox{$j<h$;} \\
    (\pi^j\models p_m)=(\pi^{j+1}\models p_{m-1}), & \hbox{$j\geq h$.}
  \end{array}
\right.
\]
Hence, for all $t\geq j$
\[
(\pi^i\models \G_{t-j}p_j)\leq \max_{H\in \mathcal{P}^{t-j}(I_t)}\bigotimes_{h\in H}(\pi^{i+h}\models \varphi).
\]
The first term corresponds to choose
$H=I_t\setminus\{h_1,\dots,h_j\}$. The converse inequality also holds,
since it derives from the monotonicity of the operation $\otimes$. Then,
passing to the limit
\[
\begin{split}
\lim_{t\rightarrow +\infty}(\pi^i\models \AG_t \varphi)&=\lim_{t\rightarrow +\infty}\max_{j\in\N}\left\{\G_{t-j} p_j\cdot \eta(j)\right\}\\
&=\max_{j\in\N}\left\{\G p_j\cdot \eta(j)\right\},
\end{split}
\]
and, indeed, we have the thesis.
\end{IEEEproof}

Note that the maximum in the definition above can be expressed
in each fuzzy logic we are considering. Indeed, in the Zadeh
Logic the maximum is simply the (standard) $\vee$, and
in a t-norm fuzzy logic it is the lattice disjunction
$\vee^w$. We decide to use the maximum
to find the best matching between the number of avoided cases, and the
penalization due to $\eta$. Indeed, if we
define the semantics of $\AG$ via the (strong) disjunction as
\[
(\pi^i\models \AG_t \varphi)=\bigoplus_{j=0}^t\bigoplus_{H\in \mathcal{P}^{t-j}(I_t)}\bigotimes_{h\in H}(\pi^{i+h}\models \varphi)\cdot \eta(j)
\]
and consider the {\L}ukasiewicz's interpretation for the connective
$\vee$, then a formula $\AG p$ will often evaluated to $1$ due to the high number
of considered cases, and (almost) independently from the evaluations
of $p$.

In the following proposition we show how to reduce the complexity of
the evaluation of operator $\AG$, by exploiting the monotonicity of the t-conorm.

\begin{proposition}\label{p:simpl-almostalways}
It is possible to evaluate the truth degree of formula $\AG_t
p$ by performing $O(n_\eta(\log(t)+1))$ comparisons, $O(t)$ applications of the norm $\otimes$, and $O(n_\eta)$ multiplications.
\end{proposition}
\begin{IEEEproof}
We consider the same technique applied in the proof of Proposition \ref{p:def-almostalways}.
Let $(a_k)_{k\leq n}$ be a finite sequence of
indices such that $\forall k\leq n$, $a_k\leq t$, and $\forall h\leq
k\leq n$, $(\pi^{a_h}\models p)\leq (\pi^{a_k}\models p)$,
then
\begin{equation}\label{e:ag}
\begin{split}
&(\pi^i\models \AG_t p)=\max_{j\leq n_\eta}\Bigg\{\!\!\!\!\!\bigotimes_{\scriptsize{\begin{array}{c}h\notin \{a_1,\dots,a_j\}\\h\leq n_\eta\end{array}}}\!\!\!\!\!(\pi^{i+h}\models p)\cdot \eta(j)\Bigg\}\\
&=\max_{1\leq j\leq n_\eta}\Bigg\{(\pi^i\models \G_t p),\!\!\!\!\!\bigotimes_{\scriptsize{\begin{array}{c}h\notin \{a_1,\dots,a_j\}\\h\leq n_\eta\end{array}}}\!\!\!\!\!(\pi^{i+h}\models p)\cdot \eta(j)\Bigg\}.
\end{split}
\end{equation}
Finding the indices $a_i$ requires at most $O(n_\eta\log(t))$
comparisons (for example applying the heapsort algorithm), and extra
$O(n_\eta)$ comparisons are used to evaluate the maximum. $O(t)$
applications of $\otimes$ are needed, observing that the operation is
associative, and, indeed, the value obtained at one step can be used
for calculating the value for the following step.
\end{IEEEproof}
From (\ref{e:ag}), we also have the following corollary.

\begin{corollary}\label{c:almostalways}
For all $\varphi\in F$ and $t\in \N$:
\[
\begin{split}
(\pi^i\models \AG_t \varphi)&\geq (\pi^i\models \G_t \varphi),\\
(\pi^i\models \AG \varphi)&\geq (\pi^i\models \G \varphi).\\
\end{split}
\]
\end{corollary}

Observe that in general it is not possible to establish a priori which
inequality holds between $(\pi^i\models \AG_t \varphi)$ and
$(\pi^i\models \AG_{t'} \varphi)$, with $t\neq t'$, as this also
depends on function $\eta$. For example, let us consider a predicate $p$
together with an avoiding function $\eta$, whose behaviors are described in
Table~\ref{tab:ag}.

\begin{table}\caption{Example of definition of a predicate $p$ and an avoiding function $\eta$.}\label{tab:ag}
\centering
\begin{tabular}{c|ccccc}
  \toprule
    $i$ & & $0$ & $1$ & $2$ & $3$\\
  \midrule
   $\pi^i\models p$ & & $0.1$ & $0.2$ & $1$ & $0.1$\\
   $\eta(i)$ & & $1$ & $0.5$ & $0.3$ & $0$\\
   \bottomrule
\end{tabular}
\end{table}
%

If we consider the Zadeh interpretation of connectives, then
$(\pi^0\models \AG_1 p)=0.1$, $(\pi^0\models \AG_2 p)=0.3$, and
$(\pi^0\models \AG_3 p)=0.06$, and the sequence $(\pi^i\models
\AG_t p)_{t\in\N}$ is not monotonic.

\subsubsection*{\bf Lasts}

Operator \lq\lq lasts\rq\rq\ ($\Lt$) is bounded, and expresses a
property that lasts for $t$ consecutive instants from now, possibly avoiding some event at the end of the considered time interval.
The semantics of this operator is defined as follows:
\[
(\pi^i\models \Lt\varphi)=\max_{0\leq j\leq\min\{t,n_\eta-1\}}\{(\pi^i\models \G_{t-j} \varphi)\cdot\eta(j)\}.
\]

\begin{proposition}\label{p:lasts}
Let $\varphi\in F$ and $t\in \N$, then the sequence $(\pi^i\models \Lt\varphi)_{t\in \N}$ is decreasing, and its limit is $(\pi^i\models \G \varphi)$. Moreover, the following inequalities hold:
\[
(\pi^i\models \G_t\varphi)\leq (\pi^i\models \Lt\varphi) \leq (\pi^i\models \AG_t\varphi).
\]
\end{proposition}
\begin{IEEEproof}
The fact that the sequence $(\pi^i\models \Lt\varphi)_{t\in \N}$ is decreasing follows immediately from the definition and from Proposition \ref{p:always}. Moreover, again from definition
\[
(\pi^i\models \G_t\varphi)\leq (\pi^i\models \Lt\varphi)\leq (\pi^i\models \G_{t-n_{\eta}}\varphi),
\]
and then passing to the limit the first part follows. The inequality $(\pi^i\models \Lt\varphi) \leq (\pi^i\models \AG_t\varphi)$ is a direct consequence of Proposition \ref{p:simpl-almostalways}.
\end{IEEEproof}

\subsubsection*{\bf Until}

The semantics of operator \lq\lq until\rq\rq\ ($\U$) and its bounded
version ($\U_t$) naturally extends the one  assigned to the
corresponding LTL operator $\mathbf{U}$, for $t>0$:
\[
\begin{split}
&(\pi^i\models \varphi\U_0 \psi)=(\pi^i\models \psi),\\
&(\pi^i\models \varphi\U_t \psi)=\max_{i\leq j\leq i+t}\left((\pi^j\models \psi)\otimes (\pi^i\models \G_{j-1}\varphi)\right),\\
&(\pi^i\models \varphi\U \psi)=\lim_{t\rightarrow +\infty}(\pi^i\models \varphi\U_t \psi),
\end{split}
\]
 Analogously to $\AG$, the maximum is used to find the best matching between the evaluation of $\psi$ and $\varphi$.

\begin{proposition}\label{p:def-until}
The semantics of operator $\U$ is well-defined. Moreover, $(\pi^i\models \varphi\U \psi)\leq (\pi^i\models \F \psi)$.
\end{proposition}
\begin{IEEEproof}
For the first part, it suffices to prove that the sequence $(\pi^i\models \varphi\U_t
\psi)_{t\in \N}$ is increasing. This is obvious as, for all $t>0$:
\[
\begin{split}
(\pi^i\models &\varphi\U_t \psi)\\&=\max\{(\pi^i\models \varphi\U_{t-1} \psi),(\pi^i\models \G_{t-1}\varphi\wedge\X^t\psi)\}.
\end{split}
\]
For the second part, let $p\in AP$ such that $\forall j\geq i$, $(\pi^j\models p)=1$. Then $(\pi^i\models \varphi\U \psi)\leq (\pi^i\models p\U \psi)$, and from Proposition \ref{p:limiti}, we have
\[
(\pi^i\models \varphi\U \psi)\leq (\pi^i\models p\U \psi)=\max_{j\geq i}(\pi^j\models \psi)\leq (\pi^i\models \F\psi).
\]
In particular, for all $t\in \N$, we can write
\begin{equation}\label{e:until}
\begin{split}
(\pi^i\models \psi)&=(\pi^i\models \varphi\U_0 \psi)\leq (\pi^i\models \varphi\U_t \psi)\\
&\leq (\pi^i\models \varphi\U \psi)\leq (\pi^i\models \F \psi).
\end{split}
\end{equation}
\end{IEEEproof}

\subsubsection*{\bf Almost until}

Operator \lq\lq almost until\rq\rq\ ($\AU$) and its bounded version
($\AU_t$) are obtained by the previous ones, by replacing operator $\G_t$ with its relaxed version $\AG_t$:
\[
\begin{split}
&(\pi^i\models \varphi\AU_0 \psi)=(\pi^i\models \psi),\\
&(\pi^i\models \varphi\AU_t \psi)=\max_{i\leq j\leq i+t}\left((\pi^j\models \psi)\otimes (\pi^i\models \AG_{j-1}\varphi)\right),\\
&(\pi^i\models \varphi\AU \psi)=\lim_{t\rightarrow +\infty}(\pi^i\models \varphi\AU_t \psi),
\end{split}
\]
for $t>0$. Similarly to $\U$, we can state the following.
\begin{proposition}\label{p:def-almostuntil}
The semantics of operator $\AU$ is well-defined. Moreover, for all $t\in \N$
\begin{equation}\label{e:almostuntil}
\begin{split}
(\pi^i\models \psi)&=(\pi^i\models \varphi\AU_0 \psi)\leq (\pi^i\models \varphi\U_t \psi)\\
&\leq (\pi^i\models \varphi\AU_t \psi)\leq (\pi^i\models \varphi\AU \psi).
\end{split}
\end{equation}
\end{proposition}
\begin{IEEEproof}
As for $\U$, we can observe that for all $t>0$,
\[
\begin{split}
(\pi^i\models& \varphi\AU_t \psi)=\\&\max\{(\pi^i\models \varphi\AU_{t-1} \psi),(\pi^i\models \AG_{t-1}\varphi\wedge\X^t\psi)\}.
\end{split}
\]
The sequence $(\pi^i\models \varphi\AU_t \psi)_{t\in \N}$ is
increasing and the semantics of $\AU$ is well-defined. The latter
part follows from Corollary \ref{c:almostalways}.
\end{IEEEproof}

Before considering further relations among operators, note that for each class of operators,
a different avoiding function can be considered. For example, we may prefer to tolerate a long delay in evaluating $\W_t$ operator, but we accept to tolerate only a few number of avoided events in evaluating $\AG_t$. In this case, we can define two functions,
$\eta_\W$ and $\eta_\G$, such that for all $i\in\N$,
$\eta_\W(i)\geq\eta_\G(i)$. However, we leave this issue for a future investigation.

As a final remark, notice that the semantics we have chosen for our operators is arbitrary,
and many other variants can be proposed. However, the properties above
show that our choice is reasonable. For example, the \lq\lq
almost \rq\rq\ operators are more lax than the traditional ones, since
their evaluation has a greater value, exactly as one would expect.

\section{Reductions and equivalences}
\label{sec:reductions}

This section prove that, under the assumption that all events
are crisp, \Fu\ reduces to LTL, and provides a set of interesting
relations between the operators of \Fu. Finally we also provide some
possible adequate set of connectives, from which it is possible to infer all the others.

\subsubsection*{\bf Reduction to LTL}

We can prove that, in some sense, the semantics of \Fu\ extends LTL,
as stated in the following proposition and theorem.

\begin{proposition}
Let $p,q\in AP$ such that for all $j\geq i$, $(\pi^j\models p), (\pi^j\models q)\in \{0,1\}$, then
\[
\begin{split}
&(\pi^i\models \F p)=1 \Leftrightarrow \pi^i\models \mathbf{F} p,\\
&(\pi^i\models \G p)=1 \Leftrightarrow \pi^i\models \mathbf{G} p,\\
&(\pi^i\models p\U q)=1 \Leftrightarrow \pi^i\models p\mathbf{U} q.
\end{split}
\]
\end{proposition}
\begin{IEEEproof}
It follows, through straightforward calculation, by applying the boundary value in Table \ref{tab:prop}.
\end{IEEEproof}

\begin{theorem}\label{t:ftltoltl}
Let for all $p\in AP$ and $i\in \N$, $\pi^i\models p\in\{0,1\}$, and $\eta(1)=0$. Then \Fu\ reduces to LTL.
\end{theorem}
\begin{IEEEproof}
First notice that, by definition, $\Soon$ reduces to $\X$, $\W_t$ to
$\F_t$, $\AG_t$ and $\Lt$ to $\G_t$, and $\AU_t$ to $\U_t$. Then, the
thesis follows by applying an argument similar to the one used in the previous proposition.
\end{IEEEproof}

\subsubsection*{\bf General relations}

\begin{figure}[]
\centering
\includegraphics[scale=.45]{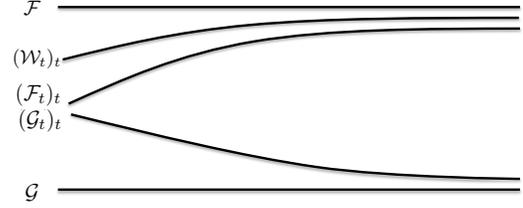}
\caption{Comparison among operators.}\label{fig:example}
\end{figure}

The relations between some of \Fu\ operators are shown in
Figure~\ref{fig:example}. Moreover, as shown in the following proposition, their values coincide only in a special case.
\begin{proposition}
Let $\varphi\in F$ and $i\in \N$, then $(\pi^i\models \F\varphi)=(\pi^i\models \G\varphi)$ if, and only if, $(\pi^j\models \varphi)$ is constant for all $j\geq i$.
\end{proposition}
\begin{IEEEproof}
For the first implication, observe that if $(\pi^j\models \varphi)$ is constant for all $j\geq i$, then for all $j,j'\geq i$, $(\pi^j\models \F\varphi)=(\pi^{j'}\models \G\varphi)=(\pi^i\models
\varphi)$. Conversely, suppose $h,k\geq i$ exist such that $(\pi^h\models
\varphi)=a<(\pi^k\models \varphi)=b$. Then from
Proposition~\ref{p:limiti} it follows that:
\[
\begin{split}
(\pi^i\models \G\varphi)&\leq \min_{j\geq i}\{\pi^j\models \varphi\}\leq a<b\\
&\leq \min_{j\geq i}(\pi^j\models \varphi)\leq (\pi^i\models \F\varphi).
\end{split}
\]
\end{IEEEproof}

\subsubsection*{\bf Adequate sets}

An adequate set of connectives for a given logic is a subset of its
connectives that is sufficient to equivalently express any
formula of the logic. For example, it is well known that $\mathbf{X}$ and $\mathbf{U}$,
together with $\wedge$ and $\neg$, form an adequate set of connectives
for LTL. Clearly, adequate sets also depend on the interpretation of
the connectives. So we denote by \Fu(Z), \Fu(G), \Fu({\L}), and \Fu($\Pi$) the
logics whose semantics is based on Zadeh, G\"{o}del-Dummett,
{\L}ukasiewicz, and Product interpretation, respectively.

Before finding an adequate sets of connectives for \Fu(Z), \Fu(G),
\Fu({\L}), and \Fu($\Pi$), we need to introduce the extra operators
$\odot^j$, for $1\leq j<n_\eta$, whose semantics is
\[
(\pi^i\models \odot^j\varphi)=(\pi^i\models \varphi)\cdot\eta(j).
\]

\begin{proposition}\label{p:fg}
Let $\varphi\in F$, then in \Fu(Z) and \Fu({\L})
\[
\G\varphi\equiv \neg\F\neg\varphi \text{ and } \F\varphi\equiv \neg\G\neg\varphi.
\]
\end{proposition}
\begin{IEEEproof}
Simply observe that, in the considered logics,
$\varphi\wedge\psi\equiv \neg(\neg \varphi\vee \neg \psi)$, and
$\varphi\vee\psi\equiv \neg(\neg \varphi\wedge \neg \psi)$.
\end{IEEEproof}

\begin{theorem}\label{t:adequate}
Let $\{\top,p_1,\dots,p_{n_\eta-1}\}\subseteq AP$, with
$\pi^i\models\top=1,\pi^i\models p_j=\eta(j)$, for all $i\in \N$, and
$1\leq j<n_\eta$. Then \Fu(Z), \Fu(G), \Fu({\L}), and \Fu($\Pi$)
admit a finite set of adequate connectives. Some of the possible adequate sets are presented in Table \ref{tab:adequate}.
\end{theorem}
\begin{IEEEproof}
It mainly follows from propositions \ref{p:within}--\ref{p:fg}, and
from the definition of the operators.
Moreover, observe that in \Fu(Z) and \Fu(G), $\F\varphi\equiv\top\U
\varphi$ and $\G\varphi\equiv\varphi\U \top$. While $\F$ and $G$ are
dual in \Fu(Z), this does not hold in $\Fu(G)$, because of the different
interpretation of the negation. Observe that in Product Logic, $\vee$
cannot be expressed in terms of $\wedge$, while this is possible in G\"{o}del-Dummett and {\L}ukasiewicz logics (see \cite{Hajek98}).
Note that the adoption of the adequate sets in Table
\ref{tab:adequate} can possibly cause a super-exponential blow-up of
the length of the formulae. For example, formula $\AG_t p$, is
equivalent in \Fu($\Pi$) to a formula of length $O(3^{2^{t+1}}\cdot
t)$ that only contains connectives
$\wedge$, $\Rightarrow$, and $\X$.
\end{IEEEproof}

\begin{table}\caption{Adequate sets for  \Fu(Z), \Fu(G), \Fu({\L}), and \Fu($\Pi$).}\label{tab:adequate}
\centering
\begin{tabular}{c|l}
  \toprule
    Logic & Adequate set\\
  \midrule
   \Fu(Z) & $\wedge,\neg,\X,\U,\AU,\odot^1,\dots,\odot^{n_\eta-1}$\\
   \Fu(G) & $\wedge,\Rightarrow,\X,\U,\AU,\odot^1,\dots,\odot^{n_\eta-1}$\\
   \Fu({\L}) & $\wedge,\Rightarrow,\X,\F,\U,\AU,\odot^1,\dots,\odot^{n_\eta-1}$\\
   \Fu($\Pi$) & $\wedge,\Rightarrow,\vee,\X,\F,\G,\U,\AU$\\
   \bottomrule
\end{tabular}
\end{table}

\section{Examples of properties and specifications}
\label{sec:example}

This section illustrates how \Fu\ can be adopted in practice to
formalize a set of properties of a smart grid. Smart grids must maximize the
availability of appliances and optimize the consumption of energy. Metering
data regarding the energy consumption are periodically
computed and are used by the Energy Management System (EMS) to balance
the work load of the appliances. In particular, the EMS sends proper
operational control data  to the appliances to schedule their tasks
and tune their functioning in order to avoid outages. To this aim, we may need to express some
statements about the amount of energy consumed and the availability of
appliances. Furthermore, we may need to tolerate a few number of
outages or some cases in which the appliances are temporarily
unavailable. Our example defines a set of formulae, under the assumption that the
smart grid controls a single
appliance ($N_1$). However, provided formulae can be easily modified
to cover the cases when more than one appliance must be controlled.

The first property, which may be necessary to evaluate, is \lq\lq $N_1$
must be available almost always during the day\rq\rq. Let
$\pi$ be the path of the daily minutes, and consider a
(fuzzy) predicate $a$ that measures whether the availability of
$N_1$ is high. More precisely, $(\pi^i\models a)$ expresses the truth
degree of proposition \lq\lq at the $i$-th minute of the day, the
availability of $N_1$ is high\rq\rq. Availability is, in general,
measured as the time difference between the instant when a request is
issued and the instant when the appliance is active. This time
difference can be estimated in seconds and this makes reasonable the
choice of minutes as time granularity.  Using this definition of
availability, we can evaluate predicate $a$ as follows. If $A_i$ is
the actual time delay of the $i$-th minute, $M_i$ the mean time delay
of the $i$-th minute of the day computed daily over the last month, and
$\sigma_i^2$ the variance, let $\Delta_i = A_i-M_i$, then
\[
(\pi_i \models a) = \left\{
  \begin{array}{ll}
    min\{1, \frac{1}{\sigma_i^2}(\Delta_i+\frac{3}{2}\sigma_i^2)\}, & \hbox{$\Delta_i \geq -\frac{3}{2}\sigma_i^2$;} \\
    0, & \hbox{otherwise.}
  \end{array}
\right.
\]

As avoiding function we can choose
$\eta(n)=e^{-(n/20)^2}$, if $n\leq 20$, and 0 otherwise.
The evaluation of formula $\G_{1440} a$ along $\pi$ gives a value corresponding,
\emph{at most}, to the worst time difference. Formula $\AG_{1440} a$,
instead, can be used when we want to tolerate the cases when the
availability of $N_1$ is fine, except for at most 20 minutes during the day. Indeed, if the availability is below the average for
no more than 4 minutes, then the evaluation of $\AG_{1440}a$ is, \emph{at least}, $e^{-(16/20)^2}\sim 0.53$, \emph{independently} from the
value of the worst minute of the day.  Observe that, if we consider the mean delay
calculated all over the day, we may obtain less expressive results, since
in case of one big delay, the evaluation of the daily availability
will dramatically decrease. 

We can also consider the crisp propositions $d$ and $c$. The former is
satisfied if new metering data are available, while the latter is
satisfied if an operational control signal is sent by the EMS to
$N_1$.  If we want to evaluate the property \lq\lq as soon as new
metering data are available, a new operational control data must be sent
by the EMS to $N_1$\rq\rq, we can formalize it as $d \Rightarrow
\W_1 c$ (or by $d \Rightarrow \Soon c$, if we do not evaluate the
formula from the first second), which allows to tolerate small delays in the
trasmission of operational control data. Instead, we cannot tolerate
small delays by using LTL, since the same proposition would be
expressed as $d \Rightarrow c$ or $d \Rightarrow \mathbf{X} c$.

Furthermore, let $s$ be a crisp proposition whose evaluation is $0$, if
the appliance is disconnected. Hence, if $p$ is the (fuzzy)
proposition ``the energy consumption is moderate'', then $\pi\models s
\U_{1440} p$ is the truth value of proposition \lq\lq there is no
outage in the day until the energy consumption is moderate\rq\rq. In
case we decide to relax our requirement to \lq\lq the outages of the day are negligible until the
energy consumption is moderate\rq\rq, we can express this requirement as $s \AU_{1440} p$. The choice of operator $\AU_{1440}$ is suitable because  $\AG_{1440} s$ allows us to neglect a few  number of
outages of the appliance during the day.

Finally, the choice of a specific interpretation for the
connectives is highly important to get more precise results, although all the inequalities we proved are still valid independently of
the interpretation. If we consider formula $\AG_{1440} s$ (\lq\lq the
daily number of outages is negligible\rq\rq), then for the evaluation
of formula $\AG_{1440} s \vee \X^{1440} \AG_{1440} s$ it is quite
natural to choose the Zadeh or G\"{o}del-Dummett interpretation,
instead of the {\L}ukasiewicz interpretation (namely, the truncated sum of
their evaluations). As a matter of fact, the predicates of this
formula do not \lq\lq saturate\rq\rq, i.e., a long sequence of days
with many outages cannot be equivalent to a day with no
outages. Still, the {\L}ukasiewicz interpretation defines a
substructural logic in which \emph{idempotency of entailment} fails, and
can be useful once we are interested in put emphasis on resource-boundedness.

\section{Conclusions}
\label{sec:conclusions}

This paper introduces \Fu, a fuzzy-time temporal logic to express
vagueness on time. The semantics of the
temporal operators provided by FTL is highly flexible, since it allows us to select a particular interpretation for the connectives, which best suits the
kind of property to be formalized. We prove that \Fu\ extends LTL, since, under the assumption that all events are crisp, \Fu\ reduces to LTL.
We show that the temporal operators introduced by our logic respect a set of
interesting relations, and we also identify adequate sets of connectives.
As future work, we are investigating a verification technique~\cite{tacas} for
checking the truth degree of the \Fu\ formulae on an automata-based
model of the system under analysis. This technique modifies the
traditional reachability analysis, according to the peculiarities of
the \Fu\ language.
Moreover, considering that FTL is particulary suitable for describing requirements of active system, in which vagueness is often embedded with uncertainty, we are planning to investigate the relationship between FTL and probabilistic languages.


\end{document}